\documentclass[11pt]{article}
\usepackage{amssymb,amsthm,amsmath,float,epsfig}
\usepackage[small,width=0.95\linewidth]{caption}
\usepackage[total={6.5in,9.0in}, top=1.0in, left=1.0in, includefoot]{geometry}
\usepackage{graphicx}
\usepackage{enumerate}
\usepackage{enumitem}
\usepackage{bm}
\usepackage{authblk}
\usepackage{setspace}
\usepackage{color}
\newtheorem*{definition}{Definition}
\newtheorem*{theorem}{Theorem}

\newcommand{\appropto}{\mathrel{\vcenter{
  \offinterlineskip\halign{\hfil$##$\cr
    \propto\cr\noalign{\kern2pt}\sim\cr\noalign{\kern-2pt}}}}}

\graphicspath{{figures/}}

\title{Control of coupled oscillator networks with application to microgrid technologies}
\author[1,2]{Per Sebastian Skardal\thanks{Electronic address: \texttt{persebastian.skardal@trincoll.edu}}}
\author[2]{Alex Arenas}
\affil[1]{Department of Mathematics, Trinity College, Hartford, CT 06106, USA}
\affil[2]{Department d'Enginyeria Inform\'{a}tica i Matem\'{a}tiques, Universitat Rovira i Virgili, 43007 Tarragona, Spain}
\date{}

\begin{document}

\maketitle

\begin{abstract}
The control of complex systems and network-coupled dynamical systems is a topic of vital theoretical importance in mathematics and physics with a wide range of applications in engineering and various other sciences. Motivated by recent research into smart grid technologies we study here control of synchronization and consider the important case of networks of coupled phase oscillators with nonlinear interactions--a paradigmatic example that has guided our understanding of self-organization for decades. We develop a method for control based on identifying and stabilizing problematic oscillators, resulting in a stable spectrum of eigenvalues, and in turn a linearly stable synchronized state. Interestingly, the amount of control, i.e., number of oscillators, required to stabilize the network is primarily dictated by the coupling strength, dynamical heterogeneity, and mean degree of the network, and depends little on the structural heterogeneity of the network itself.
\end{abstract}

\section*{Introduction}
Complex networks and complex systems describe the physical, biological, and social structures that connect our world and host the dynamical processes vital to our lives~\cite{Strogatz2001Nature,Newman2003SIAMRev,Boccaletti2006PhysRep}. The failure of such large-scale systems to operate in the desired way can thus lead to catastrophic events such as power outages~\cite{Buldyrev2010Nature,Motter2013NaturePhys}, extinctions~\cite{PaceTEE1999,Scheffer2001Nature}, and economic collapses~\cite{May2008Nature,Haldane2011Nature}. Thus, the development and design of efficient and effective control mechanisms for such systems is not only a question of theoretical interest to mathematicians, but has a wide range of important applications in physics, chemistry, biology, engineering, and the social sciences~\cite{Siljak1991,Slotine1991}. 

The roots of modern linear and nonlinear control reach back several decades, but recently research in this direction has seen a revival in physics and engineering communities. For instance, the concept of {\it structural controllability}, which is based on the paradigm of linear homogenous dynamical systems, was first introduced by Lin in \cite{Lin1974IEEE}  and more recently investigated in \cite{Liu2011Nature,Yuan2013NatComm}. These advances have enabled further progress related to structural controllability such as centrality~\cite{Liu2012PLoS}, energy~\cite{Yan2012PRL}, effect of correlations~\cite{Posfai2013SciRep}, emergence of bimodality~\cite{Jia2013NatComm}, transtion and nonlocality~\cite{Sun2013PRL}, the specific role of individual nodes~\cite{Menichetti2014PRL}, target control~\cite{Gao2014NatComm}, and control of edges in switchboard dynamics~\cite{Nepusz2012NatPhys}. Significant advances have also been made in the control of nonlinear systems, for instance the control of chaotic systems using unstable periodic orbits~\cite{Ott1990PRL}, control via pinning~\cite{Grigoriev1997PRL,Wang2002PhysA,Li2004IEEE}, control and rescue of networks using compensatory perturbations~\cite{Sahasrabudhe2011NatureComm,Cornelius2013NatureComm}, and control via structural adaptation~\cite{DeLellis2010}. Implicit in all such network control problems are the questions of (i) what form(s) of control should one choose? and (ii) how much effort is needed to attain a desired state~\cite{PasqualettiIEEE2014}?

Motivated by ongoing studies on the stability and function of power grids~\cite{Rohden2012PRL,Dorfler2013PNAS}, we study here the control of heterogeneous coupled oscillator networks~\cite{DorflerIEEE2014,Fardad2014IEEE}. Recent research into smart grid technologies has shown that certain power grid networks called {\it microgrids} evolve and can be treated as networks of Kuramoto phase oscillators~\cite{SimpsonPorco2013Automatica}. A microgrid consists of a a relatively small number of localized sources and loads that, while typically operating in connection to a larger central power grid, can disconnect itself and operate autonomously as may be necessitated by physical or economical constraints. In particular, by means of a method known as {\it frequency-drooping}, the dynamics of microgrids become equivalent to Kuramoto oscillator networks - a class of system for which a large body of literature detailing various dynamical phenomena exists~\cite{Arenas2008PhysRep}. Here we develop a control mechanism for such coupled oscillator networks, thus providing a solution with potentially direct application to the control of certain power grids.

Our goal is to induce a synchronized state, a.k.a. {\it consensus}, in a given coupled oscillator network and guarantee asymptotic stability by applying as few control gains to the network as possible. Our method is based on calculating the Jacobian of the desired synchronized state and studying its spectrum, by which we identify the oscillators in the network that contribute to unstable eigenvalues and thus destabilize the synchronized state. Importantly, our method not only identifies which oscillators require control, but also the required strength of each control gain. Interestingly, we find that the control required to stabilize a network is dictated by the coupling strength, dynamical heterogeneity, and mean degree of the network, and depends little on the structural heterogeneity of the network. In other words, the number of nodes requiring control depends surprisingly little on the network topology and degree distribution and is more sensitive to the average connectivity of the network and the dynamical parameters. While Kuramoto oscillator networks serve as our primary system of interest due both to its specific correlation with mircogrids as well as its rich body of literature, we note that our method can be applied to a much wider set of oscillator networks, provided that their linearized dynamics take a certain form. Moreover, since Kuramoto and other oscillator network models have served as a paradigmatic example for modeling and studying synchronization in various contexts, we hypothesize that our results may shed light more generally on the control of synchronization processes and could potentially give insight into other important applications such as the termination of cardiacarrhythmias~\cite{Karma2013Rev} and treatments for pathological brain dynamics~\cite{Schnitzler2005Nature}.

\section*{Results}
\subsection*{The Kuramoto model}
We consider the famous Kuramoto model for the entrainment of many coupled dissipative oscillators~\cite{Kuramoto1984}. The Kuramoto model consists of $N$ phase oscillators $\theta_i$ for $i=1,\dots,N$ that, when placed on a network dictating their pair-wise interactions, evolve according to
\begin{align}
\dot{\theta}_i=\omega_i+K\sum_{j=1}^NA_{ij}\sin(\theta_j-\theta_i).\label{eq:Kuramoto}
\end{align}
Each oscillator $i$ has a unique nature frequency $\omega_i$ that describes its preferred angular velocity in the absence of interactions, which is typically drawn randomly from a distribution $g(\omega)$. Furthermore, the global coupling strength $K$ describes the influence that oscillators have on one another via the network connectivity, which is encoded in the adjacency matrix $[A_{ij}]$. Here we focus on the simple case of an undirected, unweighted network ($A_{ij}=1$ if oscillators $i$ and $j$ are connected by a link and $A_{ij}=0$ otherwise), but we note that all results presented here easily generalize to directed and weighted networks. We also assume that the network is connected, i.e., irreducible. Over the last few decades the Kuramoto model has proven to be very useful for modeling real-world systems~\cite{Arenas2008PhysRep,Dorfler2014Automatica}, uncovering the mechanisms behind emergent collective behavior~\cite{Strogatz2003,Ott2008Chaos}, exploring additional effects such as time-delays~\cite{Lee2009PRL} and community structure~\cite{Skardal2012PRE}, and finding optimal networks structure~\cite{Skardal2014PRL}.

\begin{figure}[t]
\centering
\epsfig{file =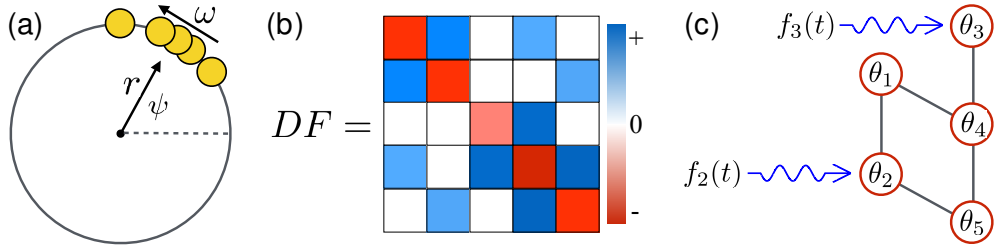, clip =,width=0.8\linewidth }
\caption{{\bf Control of coupled oscillator networks}. For a five-oscillator network, illustration of (a) the fully synchronized state with the Kuramoto order parameter, (b) the Jacobian of a stable synchronized state, and (c) control applied to oscillators $i=2$ and $3$.} \label{fig1}
\end{figure}

Depending on the coupling strength $K$, as well as the frequency vector $\bm{\omega}$ and the network topology, the steady-state dynamics of Equation~(\ref{eq:Kuramoto}) can attain many different states that included complete incoherence, partial synchronization, and full synchronization. The latter is characterized by $\lim_{t\to\infty}|\dot{\theta}_j(t)-\dot{\theta}_i(t)|=0$ and is also referred to as full phase-locking, frequency-synchronization, or consensus. The fully synchronized state (henceforth called the synchronized state) typically displays a large degree of phase synchronization $r\approx1$, where $re^{i\psi}=N^{-1}\sum_{j=1}^Ne^{i\theta_j}$ is the standard Kuramoto order parameter. In Figure~\ref{fig1}(a) we illustrate a synchronized state in a group of five oscillators, each moving with an angular velocity of $\omega$. The order parameter is illustrated as vector of length $r$ with an offset angle $\psi$ from the positive real axis.

\subsection*{Stability and control}

In this Article we address the problem of control by first assuming that due to the system parameters (i.e., the coupling strength, natural frequency sequence, or network topology), the steady-state dynamics of Equation~(\ref{eq:Kuramoto}) are at least partially incoherent, i.e., one or more oscillators remains desynchronized. In the example of the power grid, a single desynchronized oscillator represents a single power failure, but can have further damaging effects, in particular triggering a cascade of additional failures and ultimately a power outage~\cite{Motter2013NaturePhys}. Thus, our goal is to find a synchronized state and stabilize it. If all oscillators are initially synchronized, then our goal is trivially realized, however our method can be used to make this state more robustly stable. In a synchronized state like the one we seek here we expect the oscillators to be clustered in a single reasonably tight cluster such that $|\theta_j-\theta_i|\ll1$, and thus Equation~(\ref{eq:Kuramoto}) can be linearized to
\begin{align}
\dot{\theta}_i\approx\omega_i-K\sum_{j=1}^NL_{ij}\theta_j,\label{eq:Linear}
\end{align}
where $L$ is the network Laplacian matrix with entries $L_{ij}=\delta_{ij}\sum_lA_{il}-A_{ij}$ and $\delta_{ij}$ is the Kronecker delta. A straight-forward analysis yields a ``target'' synchronized state (within the rotating reference frame $\theta\mapsto\theta+\langle\omega\rangle t$) given by the vector $\bm{\theta}^*=K^{-1}L^\dagger\bm{\omega}$, where $L^\dagger$ is the pseudo-inverse of the Laplacian~\cite{BenIsrael1974}. (We summarize a derivation of this result in Materials and Methods.) We note that, since the system is assumed to be partially incoherent, the fixed point $\bm{\theta}=\bm{\theta}^*$ either does not exist or is unstable. However, we take $\bm{\theta}=\bm{\theta}^*$ to represent the closest synchronized fixed point for the given parameter values, and therefore we use it a target. We also note that, while Equation~(\ref{eq:Linear}) was directly obtained from linearizing Equation~(\ref{eq:Kuramoto}), other systems of more general forms yield equivalent linearizations and therefore can also be controlled using the method we provide here. We present an example of such a general system with arbitrary coupling function (e.g., see \cite{Skardal2015PRE}) in the Materials and Methods.

The stability of $\bm{\theta}=\bm{\theta}^*$ is dictated by the Jacobian matrix whose entries are defined $[DF]_{ij}=\partial\dot{\theta}_i/\partial\theta_j$, and is stable if all the eigenvalues of $DF|_{\bm{\theta}^*}$ are non-positive. In our case we have that
\begin{align}
DF_{ij}=\left\{\begin{array}{rl} -K\sum_{j\ne i}A_{ij}\cos(\theta_j^*-\theta_i^*)&\text{if }i=j \\ KA_{ij}\cos(\theta_j^*-\theta_i^*)&\text{otherwise.}\end{array}\right.\label{eq:DF}
\end{align}
We note that each row (and column) of $DF$ sums to zero, i.e., satisfies $DF_{ii}=-\sum_{j\ne i}DF_{ij}$. This is a particularly convenient property for using the Gershgorin circle theorem~\cite{Golub}, which implies that eigenvalues of $DF$ lie within the union of closed discs $D_i$ for $i=1,\dots,N$, which are each centered at $DF_{ii}$ and have radius $R_i$, where $R_i=\sum_{j\ne i}|DF_{ij}|$. (The full theorem is given in Materials and Methods.) In particular, if all the off-diagonal entries of $DF$ are non-negative, then it follows that each Gershgorin disc is contained in the left-half plane, implying that all eigenvalues are non-positive and the solution is stable. An illustration of this case is presented in Figure~\ref{fig1}(b). If, however, one or more non-diagonal entries of $DF$ are negative, then each Gershgorin disc corresponding to a row with a negative off-diagonal entry enters the right-half plane, admitting the possibility for one or more positive eigenvalues and thus destabilization. Thus, the oscillators that require control can be easily identified as those whose corresponding rows have one or more negative off-diagonal entries.

We aim to stabilize the synchronized solution by adding one or more control gains to the system, as illustrated in Figure~\ref{fig1}(c). In following recent literature, we will refer to oscillators to which we apply control as {\it driver nodes}, and to oscillators to which we do not apply control as {\it free nodes}. We choose the control gains to take the form $f_i(t) = F_i\sin(\phi_i-\theta_i)$, where $F_i$ is the strength of the $i^{th}$ control gain and $\phi_i$ is a target phase that can in principle depend on either local or global information, and vary in time. Here we focus on the choice of target phase $\phi_i=\theta_i^*$, and discuss other possibilities below. Since the control gain depends on the current state of the system, this can be though of as a form of feedback control. The new dynamics are then given by
\begin{align}
\dot{\theta}_i=\omega_i+K\sum_{j=1}^NA_{ij}\sin(\theta_j-\theta_i)+F_i\sin(\theta_i^*-\theta_i),\label{eq:KuramotoControl}
\end{align}
where we take $F_i=0$ for free nodes. While the off-diagonal entries of $DF$ remain unaltered, the new diagonal entries are given by $DF_{ii}=-K\sum_{j\ne i}A_{ij}\cos(\theta_j^*-\theta_i^*)-F_i$. Thus, we set coupling gain strength of each driver node $i$ such that it satisfies $F_i\ge K\sum_{j\ne i}A_{ij}[|\cos(\theta_j^*-\theta_i^*)|-\cos(\theta_j^*-\theta_i^*)]$. This ensures that all Gershgorin discs are contained in the left-half plane, implying that (up to the linear approximation of $\bm{\theta}^*$) all eigenvalues are non-positive and the synchronized state is stable. (In the case of a directed network this implies that all eigenvalues have non-positive real part and the synchronized state is stable.)

We now briefly comment on the choice of target phases $\phi_i$ in the control gains. In the method outlined above, we have set the target phase equal to the steady-state phase, $\phi_i=\theta_i^*$. This is a convenient choice for the derivation described above. Additionally, we find that in practice other choices also yield positive results. In particular, one choice that tends to yield slightly better results is to force each driver node towards the center of the synchronized cluster, i.e., $\phi_i=\overline{\phi}$, where we assume the cluster is centered at the angle $\overline{\phi}$. Target phases can also be chosen according to global or distributed control strategies. In particular, given the standard Kuramoto order parameter $re^{i\psi}=\sum_{j}e^{i\theta_j}/N$ or the set of local order parameters $r_ie^{i\psi_i}=\sum_jA_{ij}e^{i\theta_j}$, the choices $\phi_i=\psi$ and $\phi_i=\psi_i$ correspond to global and distributed control strategies that typically yield favorable results. We also note before presenting examples that since the method outlined above depends on the approximation of the steady-state solution $\bm{\theta}^*\approx K^{-1}L^\dagger\bm{\omega}$, in practice we add a buffer margin when identifying unstable oscillators, looking for non-diagonal entries of $DF$ that are not necessarily negative, but rather less than some $\epsilon>0$. We find that the choice $\epsilon=0.2$ is sufficient, and is what we use in the examples below.

\subsection*{Control of random networks}

\begin{figure}[t]
\centering
\epsfig{file =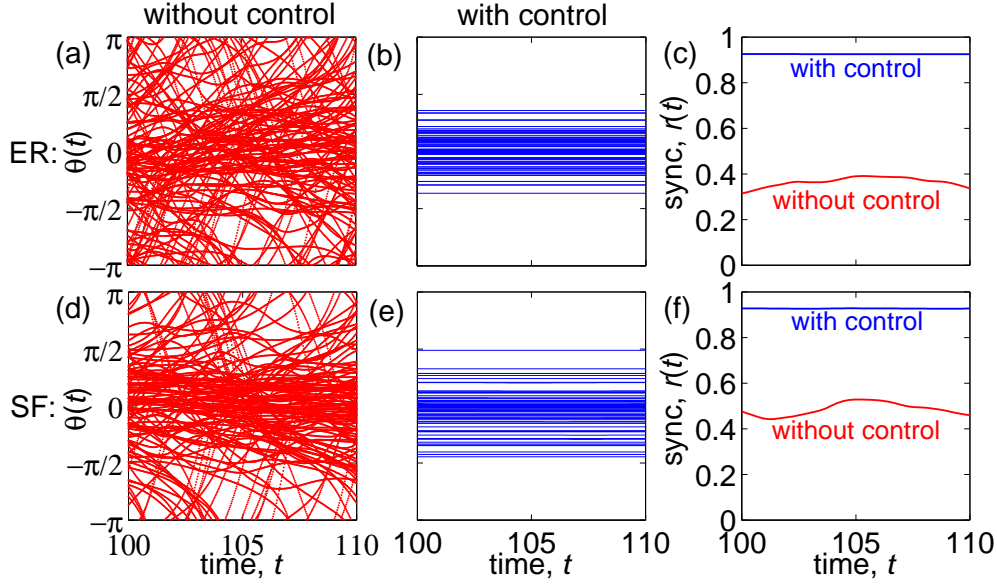, clip =,width=0.8\linewidth }
\caption{{\bf Control of coupled oscillator networks: random networks}. Example of control applied to a coupled oscillator network with ER and SF topologies (top and bottom rows, respectively) with $\langle k\rangle=6$ and $N=1000$. The coupling strength is $K=0.4$. Time series for phases $\theta_i(t)$ of $10\%$ of oscillators (a), (d) without and (b), (e) with control. Driver nodes constituted $37.1\%$ and $44.5\%$ of the ER and SF networks, respectively. (c), (f) Time series for the degree of phase synchronization $r(t)$ for the networks without (red) and with (blue) control.} \label{fig2}
\end{figure}

We now demonstrate our approach by considering two types of random networks: Erd\H{o}s-R\'{e}nyi~\cite{Erdos1960} (ER) networks and scale-free (SF) networks. Each ER network is constructed using a fixed link probability $p$ and each SF network is built using the configuration model~\cite{Molloy1995} with a degree sequence drawn from the distribution $P(k)\propto k^{-\gamma}$ with $\gamma=3$ and enforced minimum degree $k_0$. To tune the mean degree $\langle k\rangle$ of each network we set either $p=\langle k\rangle/(N-1)$ or $k_0=\langle k\rangle/(\gamma-1)$. Figure~\ref{fig2} illustrates our results with an example of each type of network, where we have used networks of size $N=1000$ with mean degree $\langle k\rangle=6$, set the coupling strength $K=0.4$, and used natural frequencies drawn from a uniform distribution with zero mean and unit variance. The top and bottom rows display the results for the ER and SF networks, respectively, displaying the time series $\theta(t)$ of a randomly selected $10\%$ of the oscillators without control [first column, panels (a) and (d)] and with control [middle column, panels (b) and (e)], after discarding a long transient. The difference between no control to control is quite drastic, with a large fraction of desynchronized oscillators without control, and full synchronization with control. Driver nodes constituted $37.1\%$ and $44.5\%$ of the ER and SF networks, respectively, to attain the synchronized state. In the last column [panels (c) and (f)] we present the degree of phase synchronization, plotting the order parameter $r(t)$ for both solutions with and without control. Unlike the solution without control, which fluctuates significantly at a relative low value, the solution with control reaches a steady value near $r(t)\approx1$.

\begin{figure}[t]
\centering
\epsfig{file =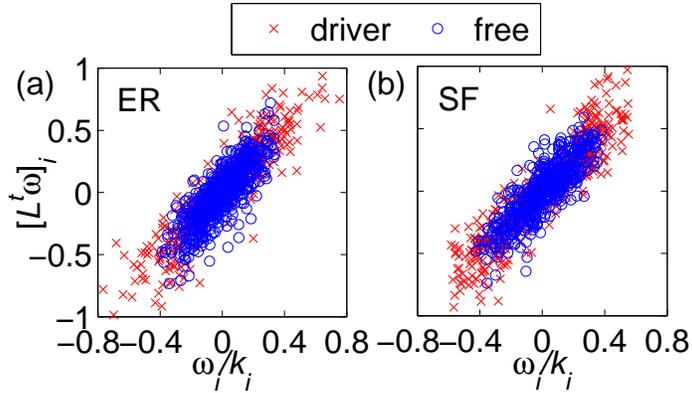, clip =,width=0.55\linewidth }
\caption{{\bf Driver and free nodes.} $[L^\dagger\bm{\omega}]_i$ vs $\omega_i/k_i$ for driver (red crosses) and free (blue circles) nodes in the (a) ER and (b) SF networks used in Figure~\ref{fig2}.} \label{fig3}
\end{figure}

Next we investigate the properties of driver and free nodes by revisiting our method. This is an essential question due to the heterogeneity of the oscillators both in terms of network structure (i.e., degree distribution) and also local dynamics (i.e., natural frequencies). An oscillator $i$ is a driver node if one of its off-diagonal entries $DF_{ij}\propto\cos(\theta_j^*-\theta_i^*)$ is negative, and therefore oscillators with large (small) steady-state values $\theta_i^*\appropto |[L^\dagger\omega]_i|$ tend to be driver (free) nodes. Furthermore, we find that these values scale approximately linearly with the ratio of the natural frequencies to degrees, i.e., $[L^\dagger\omega]_i\appropto\omega_i/k_i$. We illustrate this is Figure~\ref{fig3}, where we plot the relationship $[L^\dagger\omega]_i$ vs $\omega_i/k_i$ for the example ER and SF networks presented above [panels (a) and (b), respectively], denoting driver nodes with red crosses and free nodes with blue circles. These results show the important role that dynamics, in addition to network structure, plays in dictating controlling the system. In particular, driver nodes of the system tend to balance a large ratio (in absolute value) of natural frequencies to degrees.

\begin{figure}[t]
\centering
\epsfig{file =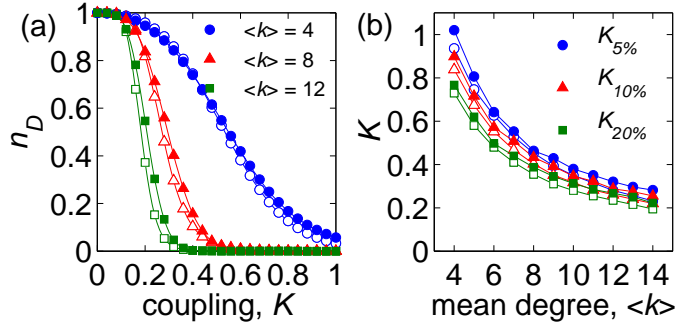, clip =,width=0.55\linewidth }
\caption{{\bf Fraction of driver nodes.} (a) The fraction of driver nodes $n_D$ in ER (unfilled symbols) and SF (filled symbols) as a function of coupling strength $K$ for mean degrees $\langle k\rangle=4$, $8$, and $12$ (blue circles, red triangle, and green squares, respectively). (b) The required coupling strengths $K_{5\%}$, $K_{10\%}$, and $K_{20\%}$ required to achieve consensus given $n_D=0.05$, $0.1$, and $0.2$ (blue circles, red triangle, and green squares, respectively). Each data point is the average over 100 network realizations of size $N=1000$, each averaged over $100$ natural frequency realizations.} \label{fig4}
\end{figure}

Finally, we quantify the overall effort required for consensus by studying how the fraction of driver nodes, denoted $n_D=N_D/N$, where $N_D$ is the total number of driver nodes, depends on both the system's dynamical and structural parameters. Presenting our results in Figure~\ref{fig4}, we first explore how the fraction of driver nodes depends on the coupling strength by plotting in panel (a) $n_D$ vs $K$ for both ER and SF networks with mean degrees $\langle k\rangle=4$, $8$, and $12$ (blue circles, red triangles, and green squares, respectively). Results for ER and SF networks are plotted with unfilled and filled symbols, respectively, and each curve represents an average over $100$ network realizations, each averaged over $100$ random natural frequency realizations. While it is expected that $n_D$ decreases monotonically with $K$, the curves' dependence on network topology and mean degree is nontrivial. In particular, the the shape of $n_D$ vs $K$ depends more sensitively on the mean degree than the topology, suggesting that network heterogeneity has little effect on overall control in comparison to average connectivity. In light of the significant dependence of overall control on the coupling strength, we investigate the coupling strength required to synchronize a network if limited mount of control is available. To this end we calculate for each family of networks the required coupling strengths $K_{5\%}$, $K_{10\%}$, and $K_{20\%}$ for which, on average, a fraction $n_D=0.05$, $0.1$, and $0.2$ will achieve synchronization as a function of the average degree $\langle k\rangle$. We plot the results in Figure~\ref{fig4}~(b). We point out again that ER and SF networks behave very similarly on average, and that with a larger mean degree, a smaller coupling strength is required to achieve synchronization.

\section*{Discussion}

Theoretical and practical aspects of the control of dynamical processes  remains an important and ongoing area of interdisciplinary research at the intersection between mathematics, physics, biology, chemistry, engineering, and the social sciences. Control of complex networks and complex systems is particularly important since together they comprise most of the world we live in~\cite{Meyn2008}, however the nonlinear nature of realistic dynamical processes and the complex network topologies of real networks represent challenges for the scientific community. Building on concepts from classical linear control theory~\cite{Lin1974IEEE}, recent work has made significant advances in understanding structural controllability~\cite{Liu2011Nature,Yuan2013NatComm}, and significant progress has been made in the development of control mechanisms for networks of nonlinear systems~\cite{Ott1990PRL,Grigoriev1997PRL,Cornelius2013NatureComm}. Nonetheless, due to the problem-sensitive nature of most real-world problems and applications requiring control techniques, further progress in designing and implementing efficient and effective control mechanisms for a wide range of problems with practical constraints remains an important avenue of research.

In this Article we have focused on the control of synchronization, i.e., consensus, in coupled oscillator networks. 
Our primary inspiration has been advances in the research of power grid networks~\cite{Gellings2004PhysicsToday,Dorfler2012SIAMControl}. In particular, recent studies have shown that certain power grids known as microgrids can be treated as Kuramoto oscillator networks~\cite{SimpsonPorco2013Automatica,Kuramoto1984}. Here we have presented a control method that can easily be applied to Kuramoto networks and other phase oscillator networks, thus providing a control framework with potentially direct application to these new technologies. Our method is based on identifying and stabilizing a synchronized state for a given network via spectral properties of the Jacobian matrix and we have demonstrated its effectiveness on both Erd\H{o}s-R\'{e}nyi and scale-free networks. We have observed that driver nodes, i.e., oscillators that require control, tend to balance (in absolute value) large natural frequencies with small degrees. Furthermore, the overall amount of control required to achieve synchronization decreases with both coupling strength and mean degree, while the total effort required to attain a synchronized state depends sensitively on the average connectivity of the network and the dynamical parameters, but surprisingly little on the network topology and degree distribution. These results enhance our understanding of and ability to understand, optimize, and ultimately control synchronization in power-grid networks (see in particular ~\cite{Motter2013NaturePhys,Dorfler2013PNAS}), and more generally complement important work on the control of network-coupled nonlinear dynamical systems~\cite{Li2004IEEE,Cornelius2013NatureComm,DeLellis2010}.

While our central inspiration and target application is in the area of power grid technology, synchronization phenomena plays a vital roll in a variety of complex processes that occur in both natural and man-made systems, including healthy cardiac behavior~\cite{Karma2007PhysicsToday}, functionality of cell circuits~\cite{Prindle2011Nature}, stability of pedestrian bridges~\cite{Strogatz2005Nature}, and communications security~\cite{Cuomo1993PRL}. Given this broad range of applications, we hypothesize that our findings here may potentially shed some light on the control of synchronization in other contexts, for instance cardiac physiology and neuroscience. For instance, a large amount of research has recently been devoted to the development of cardiac arrhythmia treatments that require minimal shock to knock out fatal asynchronous behavior such as cardiac fibrillation~\cite{Luther2011Nature} and the promotion of normal brain oscillations~\cite{Gross2004PNAS} while repressing disorders such as Parkinson's disease which are associated with abnormal oscillations~\cite{Hammond2007Trends}.

\section*{Materials and Methods}

\subsection*{Steady state solution}
To derive the steady state solution $\bm{\theta}^*=K^{-1}L^\dagger\bm{\omega}$, we begin with Equation~(\ref{eq:Linear}), which represents the linearized dynamics of Equation~(\ref{eq:Kuramoto}). Recall that this linearization requires that we are searching for a synchronized state where all oscillators are tightly packed in a single cluster so we expect that $|\theta_j-\theta_i|\ll1$. We also note that the mean frequency of all oscillators is given by the mean natural frequency $\langle\omega\rangle$. For simplicity we enter the rotating frame $\theta\mapsto\theta+\langle\omega\rangle t$, effectively setting the mean frequency to zero. It is then convenient to write Equation~(\ref{eq:Linear}) in vector form, i.e.,
\begin{align}
\dot{\bm{\theta}}\approx\bm{\omega}-KL\bm{\theta},\label{eq:linearVector}
\end{align}
where $L$ is the network Laplacian whose entries are defined $L_{ij} = \delta_{ij}\sum_{l}A_{il}-A_{ij}$. While $L$ has a zero eigenvalue, denoted $\lambda_1=0$, rendering it non-invertible, it does have a pseudo-inverse defined using its other eigenvalues (which are non-zero provided that the network is connect) and corresponding eigenvectors, $L^\dagger=\sum_{j=2}^N\lambda_j^{-1}\bm{v}^j\bm{v}^{jT}$~\cite{BenIsrael1974}. Each eigenvector is normalized such that $\{\bm{v}^j\}_{j=2}^N$ forms an orthonormal basis for the space of vectors in $\mathbb{R}^N$ with zero mean. Thus, both $L$ and $L^\dagger$ share a nullspace which is spanned by the eigenvector $\bm{v}^1\propto\bm{1}$, and therefore map vectors onto the space of zero-mean vectors in $\mathbb{R}^N$. With the pseudoinverse in hand, we can finally obtain the desired steady-state solution by setting $\dot{\bm{\theta}}=\bm{0}$ and solving for $\bm{\theta}$, which yields the solution $\bm{\theta}^*=K^{-1}L^\dagger\bm{\omega}$, as desired.

\subsection*{General oscillator networks}
Here we present an example of a more general oscillator network than that in Equation~(\ref{eq:Kuramoto}) that can be controlled using the same method detailed above. In particular, we generalize to account for an arbitrary coupling function $H(\theta)$, yielding
\begin{align}
\dot{\theta}_i=\omega_i+K\sum_{j=1}^NA_{ij}H(\theta_j-\theta_i).\label{eq:General01}
\end{align}
We assume that $H(\theta)$ is $2\pi$-periodic and at least once continuously differentiable. Importantly, $H$ need not satisfy $H(0)=0$, and thus coupling between neighboring oscillators can be {\it frustrated}~\cite{Skardal2015PRE}, denoting that even when two oscillators are exactly equal, their interaction term does not vanish. Provided that the coupling frustration is not too large, e.g., $H(0)/\sqrt{2}H'(0)\ll1$, a tightly clustered synchronized state is attainable, and linearizing Equation~(\ref{eq:General01}) yields
\begin{align}
\dot{\theta}_i\approx\omega_i +KH(0)k_i-KH'(0)\sum_{j=1}^NL_{ij}\theta_j.\label{eq:General02}
\end{align}
Importantly, by defining the quantities $\tilde{\omega}_i=\omega_i+KH(0)k_i$ and $\tilde{K}=KH'(0)$, it is easy to see that the linearized dynamics of Equation~(\ref{eq:General02}) are of the same form as Equation~(\ref{eq:Linear}), and therefore the control method we present above can be readily applied.

\subsection*{Gershgorin circle theorem}
\begin{definition}
(Gershgorin Discs) Let $M$ be an $N\times N$ complex matrix. For $i=1,\dots,N$ let $R_i=\sum_{j\ne i}|M_{ij}|$ be the sum of absolute values of non-diagonal elements of row $i$, and define $D(M_{ii},R_i)$ closed disc of radius $R_i$ centered at $M_{ii}$. $D_i=D(M_{ii},R_i)$ is the $i^{th}$ Gershgorin Disc.
\end{definition}
\begin{theorem}
(Gershgorin) All eigenvalues of the matrix $M$ lie within the union $\cup_{i=1}^ND_i$ of Gershgorin discs.
\end{theorem}

\bibliographystyle{plain}

\begin{thebibliography}{99}
\bibitem{Strogatz2001Nature} S. H. Strogatz, Exploring complex networks. {\it Nature} {\bf 410}, 268--276 (2001).
\bibitem{Newman2003SIAMRev} M. E. J. Newman, The structure and function of complex networks. {\it SIAM Rev.} {\bf 45}, 167--256 (2003).
\bibitem{Boccaletti2006PhysRep} S. Boccaletti, V. Latora, Y. Moreno, M. Chavez, D.--U. Hwang, Complex networks: structure and dynamics. {\it Phys. Rep.} {\bf 424}, 175--308 (2006).
\bibitem{Buldyrev2010Nature} S. V. Buldyrev, R. Parshani, G. Paul, H. E. Stanley, S. Havlin, Catastrophic cascade failures in interdependent networks. {\it Nature} {\bf 464}, 1025--1028 (2010).
\bibitem{Motter2013NaturePhys} A. E. Motter, S. A. Myers, M. Anghel, T. Nishikawa, Spontaneous synchrony in power grid networks. {\it Nature Phys.} {\bf 9}, 191--197 (2013).
\bibitem{PaceTEE1999} M. L. Pace, J. J. Cole, S. R. Carpenter, J. F. Kitchell, Trophic cascades revealed in diverse ecosystems. {\it Trens Ecol. Evol.} {\bf 14}, 483--488 (1999).
\bibitem{Scheffer2001Nature} M. Scheffer, S. Carpenter, J. A. Foley, C. Folke, B. Walker, Catastrophic shifts in ecosystems. {\it Nature} {\bf 413}, 591--596 (2001).
\bibitem{May2008Nature} R. M. May, S. A. Levin, G. Sugihara, Complex systems: ecology for bankers. {\it Nature} {\bf 451}, 893--895 (2008).
\bibitem{Haldane2011Nature} A. G. Haldane, R. M. May, Systemic risk in banking ecosystems. {\it Nature} {\bf 469}, 351--355 (2011).
\bibitem{Siljak1991} D. D. \v Siljak, {\it Decentralized Control of Complex Systems} (Academic Press, Boston, 1991).
\bibitem{Slotine1991} J.--J. Slotine, W. Li, {\it Applied Nonlinear Control} (Prentice-Hall,1991).
\bibitem{Lin1974IEEE} C.--T. Lin, Structural Controllability. {\it IEEE Trans. Autom. Control} {\bf 19}, 201--208 (1974).
\bibitem{Liu2011Nature} Y.--Y. Liu, J.--J. Slotine, A.--L. Barab\'{a}si, Controllability of complex networks. {\it Nature} {\bf 473}, 167--176 (2011).
\bibitem{Yuan2013NatComm} Z. Yuan, C. Zhao, Z. Di, W.--X. Wang, Y.--C. Lai, Exact controllability of complex networks. {\it Nat. Commun.} {\bf 4}, 2447 (2013).
\bibitem{Liu2012PLoS} Y.--Y. Liu, J.--J. Slotine, A.--L. Barab\'{a}si, Control centrality and hierarchical structure in complex networks. {\it PLoS ONE} {\bf 7}, e444459 (2012).
\bibitem{Yan2012PRL} G. Yan, Y.--C. Lai, C.--H. Lai, B. Li, Controlling complex networks: How much energy is needed? {\it Phys. Rev. Lett.} {\bf 108}, 218703 (2012).
\bibitem{Posfai2013SciRep} M. P\'{o}sfai, Y. Y. Liu, J.--J. Slotine, A.--L. Barab\'{a}si, Effect of correlations on network controllability. {Sci. Rep.} {\bf 3}, 1067 (2013).
\bibitem{Jia2013NatComm} T. Jia, Y.--Y. Liu, E. Cs\'{o}ka, M. P\'{o}sfai, J.--J. Slotine, A.--L. Barab\'{a}si, Emergence of bimodality in controlling complex networks. {\it Nat. Commun.} {\bf 4}, 2002 (2013).
\bibitem{Sun2013PRL} J. Sun, A. E. Motter, Controllability transition and nonlocality in network control. {\it Phys. Rev. Lett.} {\bf 110}, 208701 (2013).
\bibitem{Menichetti2014PRL} G. Menichetti, L. Dall'Asta, G. Bianconi, Network controllability is determined by the density of low in-degree and out-degree nodes. {\it Phys. Rev. Lett.} {\bf 113}, 078701 (2014).
\bibitem{Gao2014NatComm} J. Gao, Y.--Y. Liu, R. M. D'Souza, A.--L. Barab\'{a}si, Target control of complex networks. {\it Nat. Commun.} {\bf 5}, 5415 (2014).
\bibitem{Nepusz2012NatPhys} T. Nepusz, T. Vicsek, Controlling edge dynamics in complex networks. {\it Nature Phys.} {\bf 8}, 568--573 (2012).
\bibitem{Ott1990PRL} E. Ott, C. Grebogi, J. A. Yorke, Controlling Chaos. {\it Phys. Rev. Lett.} {\bf 64}, 1196--1199 (1990).
\bibitem{Grigoriev1997PRL} R. O. Grigoriev, M. C. Cross, H. G. Schuster, Pinning control of spatiotemporal chaos. {\it Phys. Rev. Lett.} {\bf 79}, 2795--2798 (1997).
\bibitem{Wang2002PhysA} X. F. Wang, G. Chen, Pinning control of scale-free dynamical networks. {\it Phys. A} {\bf 310}, 521--531 (2002).
\bibitem{Li2004IEEE} X. Li, X. Wang, G. Chen, Pinning a complex dynamical network to its equilibrium. {\it IEEE Trans. Circuits Syst., I: Fundam. Theory Appl.} {\bf 51}, 2074--2087 (2004).
\bibitem{Sahasrabudhe2011NatureComm} S. Sahasrabudhe, A. E. Motter, Rescuing ecosystems from extinction cascades through compensatory perturbations. {\it Nat. Commun.} {\bf 2}, 170 (2011).
\bibitem{Cornelius2013NatureComm} S. P. Cornelius, W. L. Kath, A. E. Motter, Realistic control of network dynamics. {\it Nat. Commun.} {\bf 4}, 1942 (2013).
\bibitem{DeLellis2010} P. DeLellis, M. di Bernardo, T. E. Gorochowski, G. Russo, Synchronization and control of complex networks via contraction, adaption and evolution. {\it Circuits Syst. Mag.} {\bf 10}, 64--82 (2010).
\bibitem{PasqualettiIEEE2014} F. Pasqualetti, S. Zampieri, F. Bullo, Controllability metrics, limitations and algorithms for complex networks. {\it IEEE Trans. Control of Netw. Syst.} {\bf 1}, 40--52 (2014).
\bibitem{Rohden2012PRL} M. Rohden, A. Sorge, M. Timme, D. Witthaut, Self-organized synchronization in decentralized power grids. {\it Phys. Rev. Lett.} {\bf 109}, 064101 (2012).
\bibitem{Dorfler2013PNAS} F. D\"{o}rfler, M. Chertkov, F. Bullo, Synchronization in complex oscillator networks and smart grids. {\it Proc. Natl. Acad. Sci. U.S.A.} {\bf 110}, 2005--2010 (2013).
\bibitem{DorflerIEEE2014} F. D\"{o}rfler, M. R. Jovanovi\'{c}, M. Chertkov, F. Bullo, Sparsity-promoting optimal wide-area control of power networks. {\it IEEE Trans. Power Syst.} {\bf 29}, 2281--2291 (2014).
\bibitem{Fardad2014IEEE} M. Fardad, F. Lin, M. R. Jovanovi\'{c}, Design of optimal sparse interconnection graphs for synchronization of oscillator network. {\it IEEE Trans. Autom. Control} {\bf 59}, 2457--2462 (2014).
\bibitem{SimpsonPorco2013Automatica} J. W. Simpson-Porco, F. D\"{o}rfler, F. Bullo, Synchronization and power sharing for droop-controlled inverters in islanded microgrids. {\it Automatica}, {\bf 49}, 2603--2611 (2013).
\bibitem{Arenas2008PhysRep} A. Arenas, A. D\'{i}az-Guilera, J. Kurths, Y. Moreno, C. Zhou, Synchronization in complex networks. {\it Phys. Rep.} {\bf 469}, 93--153 (2008).
\bibitem{Karma2013Rev} A. Karma, Physics of cardiac arhythmogenesis. {\it Annu. Rev. Condens. Matter Phys.} {\bf 4}, 313--337 (2013).
\bibitem{Schnitzler2005Nature} A. Schnitzler, J. Gross, Normal and pathological oscillatory communication in the brain. {\it Nat. Rev. Neurosci.} {\bf 6}, 285--296 (2005).
\bibitem{Kuramoto1984} Y. Kuramoto, {\it Chemical Oscillations, Waves, and Turbulence} (Springer, New York, 1984).
\bibitem{Dorfler2014Automatica} F. D\"{o}rfler, F. Bullo, Synchronization in complex networks of phase oscillators: A survey. {\it Automatica} {\bf 50}, 1539--1564 (2014).
\bibitem{Strogatz2003} S. H. Strogatz, {\it Sync: the Emerging Science of Spontaneous Order} (Hypernion, 2003).
\bibitem{Ott2008Chaos} E. Ott, T. M. Antonsen, Low dimensional behavior of large systems of globally coupled oscillators. {\it Chaos} {\bf 18}, 037113 (2008).
\bibitem{Lee2009PRL} W. S. Lee, E. Ott, T. M. Antonsen, Large coupled oscillator systems with heterogeneous interaction delays. {\it Phys. Rev. Lett.} {\bf 103}, 044101 (2009).
\bibitem{Skardal2012PRE} P. S. Skardal, J. G. Restrepo, Hierarchical synchrony of phase oscillators in modular networks. {\it Phys. Rev. E} {\bf 85}, 016208 (2012).
\bibitem{Skardal2014PRL} P. S. Skardal, D. Taylor, J. Sun, Optimal synchronization of complex networks. {\it Phys. Rev. Lett.} {\bf 113}, 144101 (2014).
\bibitem{BenIsrael1974} A. Ben-Israel, T. N. E. Grenville, {\it Generalized Inverses} (Springer, New York, 1974).
\bibitem{Skardal2015PRE} P. S. Skardal, D. Taylor, J. Sun, A. Arenas, Erosion of synchronization in networks of coupled oscillators. {\it Phys. Rev. E} {\bf 91}, 010802(R) (2015).
\bibitem{Golub} G. H. Golub, C. F. Van Loan, {\it Matrix Computations} (Johns Hopkins University Press, Baltimore, 1996).
\bibitem{Erdos1960} P. Erd\H{o}s, A. R\'{e}nyi, On the evolution of random graphs. {\it Publ. Math. Inst. Hung. Acad. Sci.} {\bf 5}, 17--61 (1960).
\bibitem{Molloy1995} M. Molloy, B. Reed, critical point for random graphs with a given degree sequence. {\it Random Struct. Algor.} {\bf 6}, 161--180 (1995).
\bibitem{Meyn2008} S. P. Meyn, {\it Control Techniques for Complex Networks} (Cambridge Univ. Press, 2008).
\bibitem{Gellings2004PhysicsToday} C. W. Gellings, K. E. Yeagee, Transforming the electric infrastructure. {\it Phys. Today} {\bf 57}, 45--51 (2004).
\bibitem{Dorfler2012SIAMControl} F. D\"{o}rfler, F. Bullo, Synchronization and transient stability in power networks and nonuniform Kuramoto oscillators. {\it SIAM J. Control Optim.} {\bf 50}, 1616--1642 (2012).
\bibitem{Karma2007PhysicsToday} Karma, A. \& Gilmour, R. F. Nonlinear dynamics of heart rhythm disorders. {\it Phys. Today} {\bf 60}, 51--57 (2007).
\bibitem{Prindle2011Nature} A. Prindle, P. Samayoa, I. Razinkov, T. Danino, L. S. Tsimring, J. Hasty, A sensing array of radically coupled genertic `biopixels'. {\it Nature} {\bf 481}, 39--44 (2011).
\bibitem{Strogatz2005Nature} S. H. Strogatz, D. M. Abrams, A. McRobie, B. Eckhardt, E. Ott, Theoretical mechanics: Crowd synchrony on the Millenium Bridge. {\it Nature} {\bf 438}, 43--44 (2005).
\bibitem{Cuomo1993PRL} K. M. Cuomo, A. V. Oppenheim, Circuit implementation of synchronization chaos with applications to communications. {\it Phys. Rev. Lett.} {\bf 71}, 65--68 (1993).
\bibitem{Luther2011Nature} S. Luther et al., Low-energy control of electrical turbulence in the heart. {\it Nature} {\bf 475}, 235--239 (2011).
\bibitem{Gross2004PNAS} J. Gross, F. Schmitz, I. Schnitzler, K. Kessler, K. Shapiro, B. Hommel, A. Schnitzler, Modulation of long-range neural synchrony reflects temporal limitations of visual attention in humans. {\it Proc. Natl. Acad. Sci. U.S.A.} {\bf 101}, 13050--13055 (2004).
\bibitem{Hammond2007Trends} C. Hammond, H. Bergman, P. Brown, Pathological synchronization in Parkinson's disease: networks, models and treatments. {\it Trends Neurosci.} {\bf 30}, 357--364 (2007).
\end{thebibliography}

\section*{Acknowledgements}
This work was supported by the James S. McDonnell Foundation (PSS and AA), Spanish DGICYT Grant No. FIS2012-38266 (AA), and FwET Project No. MULTIPLEX (317532) (AA).

\noindent Competing Interests: The authors declare that they have no competing interests.

\end{document}